\documentclass[prd,aps,twocolumn,a4paper,floatfix,amsmath,amssymb]{revtex4-1}
\usepackage{graphicx,color}

\begin{document}

\title{The Einstein-Vlasov system in spherical symmetry II: spherical
  perturbations of static solutions}

\author{Carsten Gundlach} 

\date{24 August 2017}

\begin{abstract}

We reduce the equations governing the spherically symmetric
perturbations of static spherically symmetric solutions of the
Einstein-Vlasov system (with either massive or massless particles) to
a single stratified wave equation $-\psi_{,tt}=H\psi$, with $H$
containing second derivatives in radius, and integrals over energy and
angular momentum. We identify an inner product with respect to which
$H$ is symmetric, and use the Ritz method to approximate the lowest
eigenvalues of $H$ numerically. For two representative background
solutions with massless particles we find a single unstable mode with
a growth rate consistent with the universal one found by Akbarian and
Choptuik in nonlinear numerical time evolutions.

\end{abstract}

\maketitle

\tableofcontents

\section{Introduction} 
\subsection{The Einstein-Vlasov system} 

The Vlasov-Einstein system describes an ensemble of particles of
identical rest mass, each of which follows a geodesic. The particles
interact with each other only through the spacetime curvature
generated by their collective stress-energy tensor, whereas particle
collisions are neglected. 

For massive particles, this is a good physical model of a stellar
cluster. For either massive or massless particles, the Einstein-Vlasov
system also serves as a well-behaved toy model of matter in general
relativity. In particular, spherically symmetric solutions of the
Einstein-Vlasov system with small data are known to exist globally in
time for massive \cite{ReinRendall1992} and massless particles
\cite{Dafermos2006}. Self-similar spherically symmetric solutions with
massless particles have been analyzed in \cite{JMMGundlach2002} and
\cite{RendallVelazquez2011}. The existence of spherically symmetric
static solutions with massive particles was proved in
\cite{ReinRendall2000}, and there is numerical evidence that at least
some are stable within spherical symmetry
\cite{AndreassonRein2006}. Spherically symmetric static solutions with
massless particles were analysed and constructed numerically in
\cite{paperI}, and we investigate their linear perturbations here. See also
\cite{AndreassonLRR2011} for a review and additional references.

\subsection{Motivation for this paper} 

This is the second paper in a series motivated by Akbarian and Choptuik's
\cite{AkbarianChoptuik} (from now on, AC) recent study of numerical
time evolutions of the massless Einstein-Vlasov system in spherical
symmetry. AC found two apparently contradictory results:

I) Taking several 1-parameter families of generic smooth initial data
and fine-tuning the parameter to the threshold of black hole
formation, AC found what is known as type-I critical collapse: in the
fine-tuning limit the time evolution goes through an intermediate
static solution. The lifetime $\Delta\tau$ of this static solution increases
with fine-tuning to the collapse threshold as 
\begin{equation}
\label{lifetimescaling}
\Delta\tau\simeq\sigma\ln|p-p_*|+{\rm const},
\end{equation}
where $p$ is the parameter of the family, $p_*$ its value at the
black-hole threshold, and $\tau$ is the proper time at the centre, in
units of the total mass of the critical
solution. (\ref{lifetimescaling}) implies the existence of a single
unstable mode growing as $\exp(\tau/\sigma)$.  AC found that $\sigma$
was approximately universal (independent of the family), with value
$\sigma\simeq 1.4\pm 0.1$, and that the metric of the intermediate
static solution was also approximately universal (up to an overall
length and mass scale). In particular, its compactness
$\Gamma:=\max(2M/r)$ was in the range $\Gamma\simeq0.80\pm 0.01$ and
its central redshift $Z_c\ge 0$ was in the range $Z_c\simeq2.45\pm
0.05$.

II) Conversely, constructing static solutions by ansatz, AC
found that these covered much larger ranges of $\Gamma$ and $Z_c$, but
that each one was at the threshold of collapse. That is, adding a
small generic perturbation to the static initial data and evolving in
time with their nonlinear code, they found that for one sign of the
perturbation the perturbed static solution collapsed while for the
other sign it dispersed. They found that $\sigma$ was in the narrow
range $\sigma\simeq1.43\pm 0.07$, compatible with the
value above.

Result II suggests that in spherical symmetry with massless particles,
the black hole-threshold coincides with the space of static
solutions. If so, then each static solution would have precisely one
unstable mode (with its sign deciding between collapse and
dispersion), with all other modes either zero modes (moving to a
neighbouring static solution) or purely oscillating. 

One aspect of Result I, namely that the spacetime of the critical
solution is universal, would imply that this universal solution has
one unstable mode (as before), but that all its other modes (including those
tangential to the black hole threshold) are decaying ones, so that the
attracting manifold of the critical solution is precisely the black
hole threshold. Indeed, this is the familiar picture of type-I
critical collapse in other matter-Einstein systems. However, this is
in apparent contradiction to Result II. 

In the first paper in this series \cite{paperI} (from now Paper I), we
used a symmetry of the {\em massless} spherically symmetric
Einstein-Vlasov system to reduce its number of independent variables
from four to three. We then numerically constructed static solutions
with compactness in the range $0.7\simeq \Gamma \leq 8/9$.  Based on
this, we conjectured that the apparent contradiction above is resolved
by the critical solution seen in fine-tuning generic initial data
being universal only to leading order, and that this leading order is
selected by the way in which it is approached during the evolution of
generic smooth initial data.

To make further progress, it seems essential to analyse the spectrum
of linear perturbations directly. This is the programme of the current
paper. In contrast to the static solutions investigated in Paper I,
their perturbations do not simplify significantly for $m=0$, and hence
all of our analysis, except for the numerical examples in
Sec.~\ref{section:numericalexamples}, will be for $m\ge 0$.

\subsection{Plan of the paper} 

In order to make the presentation self-contained and to establish
notation, we review some material from Paper I in
Sec.~\ref{section:background}. We  begin in Sec.~\ref{section:equations}
by presenting the equations of the time-dependent spherically
symmetric Einstein-Vlasov system. We do this in a form in which the
massless particle limit is regular and leads to a reduction of the
number of independent variables. We discuss static solutions in
Sec.~\ref{section:staticsolutions}, and the massless limit, for both
the time-dependent and static case, in Sec.~\ref{section:massless}.

In Sec.~\ref{section:perturbations} we then derive the spherical
perturbation equations. In Sec.~\ref{section:perturbationansatz} we
perturb the Vlasov and Einstein equations about a static background,
splitting the perturbation of the Vlasov distribution function $f$
into parts $\phi$ and $\psi$ that are even and odd, respectively under
reversing time. In Sec.~\ref{section:ztoQ} we change independent
variables from momentum to energy, as we did for the background
solutions. We quickly dispense with static perturbations in
Sec.~\ref{section:staticperturbations}, and in
Sec.~\ref{section:singlewaveequation} we reduce the perturbed Vlasov
and Einstein equations to a single integral-differential equation
$-\psi_{,tt}=H\psi$. In Sec.~\ref{section:classificationofmodes} we
dispense with the relatively trivial perturbations on regions of phase
space where the background solution is vacuum. We state the massless
limit in Sec.~\ref{section:masslessperturbations}.

In Sec.~\ref{section:perturbationspectrum} we attempt to find the
spectrum of eigenvalues. In Sec.~\ref{section:innerproduct} we
identify a positive definite inner product with respect to which $H$
is symmetric. In Sec.~\ref{section:quadraticform} we rewrite the
Hamiltonian as $H=A^\dagger A-D^\dagger D$ where $D$ is bounded,
giving us at least a lower bound on $H$. We then switch to an
approximation method, the Ritz method, which we review in
Sec.~\ref{section:Ritzmethod}. In Sec.~\ref{section:functionspace} we
specify some properties of the function space ${\Bbb V}$ in which to
look for perturbation modes, that is eigenfunctions of $H$. In
Sec.~\ref{section:numericalexamples} we pick two specific background
solutions that we obtained numerically in Paper~I and use the Ritz
method numerically. We find values of $\sigma$ in agreement with AC.

Sec.~\ref{section:conclusions} contains a summary and
outlook. Throughout the paper, $a:=b$ defines $a$, and we use units
such that $c=G=1$.

\section{Background equations} 
\label{section:background}

\subsection{Field equations in spherical symmetry} 
\label{section:equations}

We consider the Einstein-Vlasov system in spherical symmetry, with
particles of mass $m\ge 0$. We write the metric as
\begin{equation}
\label{metric}
ds^2=-\alpha^2(t,r) dt^2+a^2(t,r)dr^2
     +r^2(d\theta^2+\sin^2\theta d\varphi^2).
\end{equation}
To fix the remaining gauge freedom we set $\alpha(t,\infty)=1$. The
Einstein equations give the following equations for the first
derivatives of the metric coefficients:
\begin{eqnarray}
\frac{\alpha_{,r}}{\alpha}&=&
\frac{a^2-1}{2r}+4\pi ra^2p, \label{alphar} \\
\frac{a_{,r}}{a}&=&-\frac{a^2-1}{2r}+4\pi ra^2\rho, 
\label{ar} \\
\frac{a_{,t}}{a}&=&4\pi ra^2j,
\label{at}
\end{eqnarray}
where $p$, $\rho$ and $j$ are the radial pressure, energy density
and radial momentum density, measured by observers at constant $r$. The
fourth Einstein equation, involving the tangential pressure $p_T$, is
a combination of derivatives of these three, and is redundant modulo
stress-energy conservation.

The Vlasov density describing collisionless matter in general
relativity is defined on the mass shell $p^\mu p_\mu=-m^2$ of the
cotangent bundle of spacetime, or $f(t,x^i,p_i)$ in coordinates. We
define the square of particle angular momentum
\begin{equation}
F := p_\theta^2+\sin^2\theta\,p_\varphi^2.
\end{equation}
In spherical symmetry this is conserved, leaving to $f=f(t,r,p_r,F)$.
In order to obtain a reduction of the Einstein-Vlasov system in the
limit $m=0$, we replace $p_r$ by the new independent variable
\begin{equation}
z:={p_r\over a\sqrt{F}}.
\end{equation}
The Vlasov equation for $f(t,r,z,F)$ is
\begin{equation}
\frac{\partial f}{\partial t}
+{\alpha z\over aZ}\frac{\partial f}{\partial r}
+
\left({\alpha\over r^3 a Z}-{\alpha_{,r} Z\over a}-{z a_{,t}\over a}\right)
\frac{\partial f}{\partial z}
=0,
\label{VlasovtrzF}
\end{equation}
where we have defined the shorthand
\begin{equation}
\label{Zdefmassive}
Z(r,z,F):=\sqrt{{m^2\over F}+z^2+{1\over r^2}}={\alpha p^t\over\sqrt{F}}.
\end{equation}

The non-vanishing components of the stress-energy tensor are
\begin{eqnarray}
p:=T_r{}^r &=& \frac{\pi}{r^2} {\cal J} f\, {z^2\over Z}, \label{Trrz} \\
\rho:=-T_t{}^t &=& \frac{\pi}{r^2} {\cal J} f\, Z, \label{Tttz} \\
j:=T_t{}^r &=& - \frac{\pi}{r^2}{\alpha\over a} {\cal J} f\, z, \label{Ttrz} \\
p_T:={T_\theta}^\theta = {T_\varphi}^\varphi  &=&  \frac{\pi}{2r^4}
{\cal J} f\, {1\over Z}, \label{Tphiphiz}
\end{eqnarray}
where we have introduced the integral operator (acting to the right)
\begin{equation}
{\cal J}:=\int_0^\infty FdF \int_{-\infty}^\infty dz.
\end{equation}

\subsection{Static solutions} 
\label{section:staticsolutions}

In the static metric
\begin{equation}
\label{staticmetric}
ds^2=-\alpha_0^2(r) dt^2+a_0^2(r)dr^2
     +r^2(d\theta^2+\sin^2\theta d\varphi^2)
\end{equation}
with Killing vector $\xi:=\partial_t$, the particle energy
\begin{equation}
E:=-\xi^\mu p_\mu=- p_t=\alpha_0 \sqrt{F}Z
\end{equation}
is conserved along particle trajectories. In order to obtain a
reduction of the static Einstein-Vlasov system in the
limit $m=0$, we replace $E$ by 
\begin{equation}
\label{Qdef}
Q(r,z,F):={E^2\over F}=\alpha_0^2 Z^2.
\end{equation}
Hence we have
\begin{equation}
z=\pm \alpha_0(r)\sqrt{Q-U},
\end{equation}
where we have defined 
\begin{equation}
\label{Udef}
U(r,F):=\alpha_0^2\left({m^2\over F}+{1\over r^2}\right).
\end{equation}
The general static solution of the Vlasov equation, for simplicity
with a single potential well, is then given by
\begin{equation}
\label{kdef}
f(r,z,F)=k(Q,F).
\end{equation}
(If $U$ forms more than one potential well, $k$ could be different in each
of them \cite{Schaeffer1999}.)

The static Einstein equations are 
\begin{eqnarray}
{\alpha_0'\over \alpha_0}&=&{a_0^2-1\over2r}+{4\pi^2a_0^2\over r} {\cal I}{k}{v},
\label{staticalpham} \\
{a_0'\over a_0}&=&-{a_0^2-1\over2r}-{4\pi^2a_0^2\over r}{\cal I}{{k}\over{v}},
\label{staticam}
\end{eqnarray}
where we have defined the integral operator (acting to the right)
\begin{equation}
\label{calIdef}
{\cal I}:=\int_0^\infty FdF \int_{{U(r,F)}}^\infty \,dQ
\end{equation}
and the shorthand
\begin{equation}
\label{vdef}
{v}(r,Q,F):={|z|\over Z}={a |p^r|\over \alpha p^t}.
\end{equation}
From (\ref{Qdef}) and (\ref{Zdefmassive}) we have $dQ=2\alpha_0^2z\,dz$, and
hence on a static background, where ${\cal I}$ is defined, it is
related to ${\cal J}$ by
\begin{equation}
\label{IJrelation}
{\cal I}=\alpha_0^2{\cal J}z.
\end{equation}

The second equality in (\ref{vdef}) shows that $\pm v$ is the radial
particle velocity, expressed in units of the speed of light and
measured by static observers. (Note that by definition $v\ge 0$.) The
fact that (\ref{vdef}) can be rewritten as
\begin{equation}
\label{v2def}
v=\sqrt{1-{U\over Q}}
\end{equation}
shows that $U$ is an effective potential for the radial motion of the
particles with given constant ``energy'' $Q$ and angular momentum
squared $F$. In particular $Q=U(r,F)$ determines the radial
turning points $r$ for all particles with a given $Q$ and $F$.

We define $U_0(F)$ as the value of $U(r,F)$ at its one local maximum
$r_{0+}(F)$ in $r$. We define $r_{0-}(F)$ as the other value of $r$
for which $U$ takes the same value. We also define $U_3(F)$ as the
value of $U(r,F)$ at its one local minimum $r_{3}(F)$ in
$r$. Intuitively, $U_0(F)$ is the ``lip'' of the effective potential
for particles of a given $F$, and $U_3(F)$ its ``bottom''. We define
$U_1(F)\le U_0(F)$ as the upper boundary in $Q$ of the support of
$k(Q,F)$ and $U_2(F)\ge U_3(F)$ as the lower boundary. We define
$r_{1\pm}(F)$ and $r_{2\pm}(F)$ as the values where $Q=U_1(F)$ or
$Q=U_2(F)$.

For massless particles, particles with all values of $F$ move in the
same potential $U=U(r)$, and all other quantities we have just defined
do also not depend on $F$.  See Fig.~\ref{figure:potentialsketch} for an
illustration.

Any particles with $Q>U_0(F)$ would be unbound, and if such particles
were present in a static solution with the ansatz (\ref{kdef}) they
would be present everywhere and hence the total mass could not be
finite. We must therefore have $k(Q,F)=0$ for $Q>U_0(F)$.  
$k(Q,F)$ must either have compact support in $F$, or fall
off as $F\to\infty$ sufficiently rapidly for $\int kFdF$ to be finite.
Both assumptions are assumed implicitly in the definition
(\ref{calIdef}) of ${\cal I}$ when we formally extend the upper
integration limits in $Q$ and $F$ to $\infty$.

By contrast, real values of $z$ require $Q\ge U(r,F)$, and when
changing from $\int dz$ to $\int dQ$, the condition $Q\ge U$ must be
imposed {\em explicitly} as a limit to the integration range in
(\ref{calIdef}). Hence $k(Q,F)$ needs to be defined for all $U_3(F)\le
Q\le U_0(F)$ (it can be zero in part of that range), but not all of
that range contributes to the integral ${\cal I}$, and hence to the
stress-energy tensor, at every value of $r$.

\subsection{Reduction of the field equations for $m=0$}
\label{section:massless} 

Because $F$ is conserved, the field equations already do not contain
$\partial/\partial F$. Furthermore, $m$ and $F$ appear in the
coefficients of the field equations only in the combination
$m^2/F$. In the limit $m=0$, these coefficients become independent of
$F$. As first used in \cite{JMMGundlach2002}, the integrated Vlasov density
\begin{equation}
\bar f(t,r,z):=\int_0^\infty f(t,r,z,F)FdF
\end{equation}
then obeys the same PDE (\ref{VlasovtrzF}) as $f$ itself. The
stress-energy tensor is now given by the expressions
(\ref{Trrz}-\ref{Tphiphiz}) above, with $\bar f$ in place of $f$, and
\begin{equation}
{\bar {\cal J}}:=\int_{-\infty}^\infty dz
\end{equation}
in place of ${\cal J}$. We have reduced the Einstein-Vlasov system
to a system of integral-differential equations in the independent
variables $(t,r,z)$ only. A solution $(a,\alpha,\bar f)$ of the
reduced system represents a class of solutions $(a,\alpha,f)$, all with
the same spacetime but different matter distributions.

Similarly, in the static case with $m=0$ the Einstein
equations are given by (\ref{staticalpham}-\ref{staticam}) with
\begin{equation}
\label{barkdef}
\bar {k}(Q):=\int_0^\infty {k}(Q,F)FdF
\end{equation}
in place of $k(Q,F)$ and
\begin{equation}
{\bar{\cal I}}:= \int_{{U(r)}}^\infty \,dQ
\end{equation}
in place of ${\cal I}$.  Each solution of the reduced system
$(a_0,\alpha_0,\bar k)$ represents a class of solutions
$(a_0,\alpha_0,k)$.  We now define $U_1$ and $U_2$ as the limits of
support of $\bar k(Q)$.

In Paper I, we conjectured that the space of spherically symmetric
static solutions with massless particles is a space of single
functions of one variable, subject to certain positivity and
integrability conditions. This single function can be taken to be any
one of $\alpha_0(r)$, $a_0(r)$ or $\bar k(Q)$.

As a postscript to Paper I, we note here that a mathematically similar
result holds for Vlasov-Poisson system for a subclass of static
solutions, namely the ``isotropic'' ones, where the Vlasov function is
assumed to be a function of energy only. With the isotropic ansatz the
Vlasov density uniquely defines the Newtonian gravitational potential
and vice versa \cite{DeJonghe}. Physically, however, the two results
are different, as the relativistic result with massless particles
holds for all solutions, depending on both energy and angular
momentum.

\section{Spherical perturbation equations} 
\label{section:perturbations}
\subsection{Perturbation ansatz} 
\label{section:perturbationansatz}

In the study of the linear perturbations of a static solution we
return to the general case $m\ge 0$. We make the perturbation ansatz
\begin{eqnarray}
\alpha(t,r)&=&\alpha_0(r)+\delta\alpha(t,r), \\
a(t,r)&=&a_0(r)+\delta a(t,r), \\
\label{deltafansatz}
f(t,r,z,F)&=&{k}({Q},F) +\phi(t,r,z,F) \nonumber \\
&&+\,\psi(t,r,z,F),
\end{eqnarray}
where $(\alpha_0,a_0,k)$ is a static solution of the Einstein-Vlasov
system, and where $Q$ is given by (\ref{Qdef}). We define $\phi$ to be
even in $z$ and $\psi$ to be odd. In particular, as $\delta
f(t,r,z,F)$ must be at least once continuously differentiable ($C^1$)
to obey the Vlasov equation in the strong sense, its odd-in-$z$ part must
vanish at $z=0$, or
\begin{equation}
\label{psiboundaryz}
\psi(t,r,z=0,F)=0.
\end{equation}

We now expand to linear order in the perturbations. The linearised
Vlasov equation splits into the pair
\begin{eqnarray}
\label{phievolution}
\phi_{,t}+{L}\psi&=& 2Q k_{,Q}{{v}^2\over a_0}\delta a_{,t}, \\
\label{psievolution}
\psi_{,t}+{L}\phi &=& 2Qk_{,Q}{{v}\alpha_0\over a_0}
\left({\delta\alpha\over\alpha_0}\right)_{,r},
\end{eqnarray}
where we have defined the differential operator
\begin{equation}
{L} :={z\alpha_0\over Z a_0}{\partial\over\partial r}
+\left({\alpha_0\over
  a_0r^3Z}-{Z\alpha_0'\over a_0}\right){\partial\over\partial z}.
\end{equation}
Note, from (\ref{VlasovtrzF}), that the time-dependent Vlasov
equation in the fixed static spacetime $(a_0,\alpha_0)$ is $f_{,t}+{L}
f=0$. By construction, ${Q}$ obeys ${L} {Q}=0$, and hence ${L}
{k}({Q},F)=0$, as we have already used to construct static solutions.
The even-odd split of the perturbed Vlasov equation into the pair
(\ref{phievolution},\ref{psievolution}) reflects the fact that
reversing both $t$ and $z$ together must leave the field equations
invariant.

The perturbed Einstein equations are
\begin{eqnarray}
\left({\delta\alpha\over\alpha_0}\right)_{,r}
-\left({1\over r} +{2\alpha_0'\over\alpha_0}\right){\delta a\over a_0}
&=&{4\pi^2a_0^2\over r}{\cal J}
{z^2\over Z}\phi, \label{deltaalphaprimemJ} \\
\left({r\delta a\over a_0^3}\right)_{,r}
&=&4\pi^2{\cal J} Z\phi, \label{deltaaprimemJ} \\
\label{deltaadotmJ}
\delta a_{,t}&=&-{4\pi^2a_0^2\over r}{\cal J}z\psi.
\end{eqnarray}
We have used the background Einstein equations to eliminate all
integrals of $k$ in favour of $a_0'$ and $\alpha_0'$. The
integrability condition for $\delta a$ between (\ref{deltaaprimemJ})
and (\ref{deltaadotmJ}) is identically obyed modulo the perturbed
Vlasov equation (\ref{phievolution},\ref{psievolution}), and hence the
perturbed Einstein equation (\ref{deltaadotmJ}) is redundant [just as
  (\ref{at}) is in the nonlinear equations].

\subsection{Change of variable from $z$ to $Q$} 
\label{section:ztoQ}

To simplify the perturbation equations, we change independent
variables from $(t,r,z,F)$ to $(t,r,Q,F)$, with $Q$ again given by
(\ref{Qdef}). From the resulting transformation of partial
derivatives, we need only the following identities:
\begin{eqnarray}
\left.{\partial\over\partial t}\right|_z&=&
\left.{\partial\over\partial t}\right|_Q, \\
\label{drtransformation}
\left.{\partial\over\partial r}\right|_z&=&
\left.{\partial\over\partial r}\right|_Q+(\dots)
\left.{\partial\over\partial Q}\right|_r, \\
\label{Lr}
{L} &=&{V}\left.{\partial\over\partial r}\right|_Q,
\end{eqnarray}
where we have defined the shorthand
\begin{equation}
{V}(r,Q,F)={\alpha_0\over a_0}{v},
\end{equation}
with $v$ given by (\ref{v2def}). $V$ is the coordinate speed $dr/dt$
corresponding to the physical speed $v$.

$\partial/\partial r|_z$ on its own appears in the field equations
only acting on the metric and its perturbations, in which case the
$\partial/\partial Q|_r$ term denoted by $(\dots)$ in
(\ref{drtransformation}) is irrelevant. On the matter perturbations,
$\partial/\partial r|_z$ acts only in the combination ${L}$. From now
on $\partial/\partial r$ is understood to mean $\partial/\partial
r|_Q$.

With (\ref{Lr}) the perturbed Vlasov equation becomes
\begin{eqnarray}
\label{phiQevolution}
\phi_{,t}+V\psi_{,r}&=& 2Q k_{,Q}{{v}^2\over a_0}\delta a_{,t}, \\
\label{psiQevolution}
\psi_{,t}+V\phi_{,r} &=& 2Qk_{,Q}{{v}\alpha_0\over a_0}
\left({\delta\alpha\over\alpha_0}\right)_{,r}.
\end{eqnarray}

With (\ref{IJrelation}) and (\ref{vdef}) the perturbed Einstein
equations (\ref{deltaalphaprimemJ}-\ref{deltaadotmJ}) become
\begin{eqnarray}
\left({\delta\alpha\over\alpha_0}\right)_{,r}
-\left({1\over r} +{2\alpha_0'\over\alpha_0}\right){\delta a\over a_0}
&=&{4\pi^2a_0^2\over r\alpha_0^2}{\cal I}
{v}\phi, \label{deltaalphaprimem} \\
\left({r\delta a\over a_0^3}\right)_{,r}
&=&{4\pi^2\over \alpha_0^2}{\cal I}{\phi \over {v}}, \label{deltaaprimem} \\
\label{deltaadotm}
\delta a_{,t}&=&-{4\pi^2a_0^2\over r\alpha_0}{\cal I}\psi.
\end{eqnarray}

Equivalently, the stress-energy perturbations are given by
\begin{eqnarray}
\label{deltapstatic}
\delta p&=&{\pi\over r^2\alpha_0^2}{\cal I}v\phi, \\
\delta \rho&=&{\pi\over r^2\alpha_0^2}{\cal I}{\phi\over v}, \\
\delta j&=&-{\pi\over r^2\alpha_0 a_0}{\cal I}\psi.
\label{deltajstatic}
\end{eqnarray}

Finally, the boundary condition on $\psi$ becomes
\begin{equation}
\label{psiboundary}
\psi(t,r,U(r,F),F)=0.
\end{equation}

\subsection {Static perturbations} 
\label{section:staticperturbations}

Static perturbations can be obtained more directly by linear
perturbation of the nonlinear static equations. The infinitesimal
change $k\to k+\delta k$, $a_0\to a_0+\delta a_0$, $\alpha_0\to
\alpha_0+\delta\alpha_0$ to a neighbouring static solution gives
\begin{eqnarray}
\phi&=&\left.{d\over d\epsilon}\right|_{\epsilon=0} 
(k+\epsilon\delta k)
\left((\alpha_0+\epsilon\delta\alpha)^2Z^2,F\right) \nonumber \\ 
&=&\delta k(Q,F)+2Qk_{,Q}{\delta\alpha\over\alpha_0}, \\ 
\psi&=&0.
\end{eqnarray}
As expected, this solves the perturbed Vlasov equations
(\ref{phiQevolution}-\ref{psiQevolution}) and Einstein
equation (\ref{deltaadotm}) identically for arbitrary
functions $\delta k(Q,F)$, modulo the nontrivial perturbed Einstein equations
(\ref{deltaalphaprimem}-\ref{deltaaprimem}).

\subsection{Reduction to a single stratified wave equation} 
\label{section:singlewaveequation}

We can uniquely split any time-dependent perturbation that admits a
Fourier transform with respect to $t$ into a static part (with
frequency $\omega=0$) and a genuinely non-static part (with
frequencies $\omega\ne 0$). For genuinely non-static perturbations we
can uniquely invert $\partial/\partial t$ by dividing by $i\omega$ in
the Fourier domain.

We now substitute (\ref{deltaadotm}) into (\ref{phiQevolution}), and
(\ref{deltaalphaprimem}) and $\left({\partial\over\partial
  t}\right)^{-1}(\ref{deltaadotm})$ into (\ref{psiQevolution}), and
write the result in the compact form
\begin{equation}
\left(\begin{array}{cc}
{\partial\over\partial t} & A \\
B & {\partial\over\partial t} + \left({\partial\over\partial t}\right)^{-1}C
\end{array}\right)
\left(\begin{array}{cc}
\phi \\ \psi
\end{array}\right)=0,
\label{operatorform}
\end{equation}
where we have defined the integral-differential operators (acting to
the right)
\begin{eqnarray}
\label{Adef}
A&:=&{L} +g{v}{\cal I}, \\
B&:=&{L} -g{\cal I}{v}, \\
C&:=&hg{\cal I}
\label{Cdef}
\end{eqnarray}
with the differential operator ${L}$ now given by (\ref{Lr}), the
integral operator ${\cal I}$ defined in (\ref{calIdef}) (both acting
to the right), $v$ given by (\ref{v2def}), and the shorthand
expressions
\begin{eqnarray}
g(r,Q,F)&:=&{8\pi^2a_0\over r\alpha_0}Q\,k_{,Q}\,{v}, \\
h(r)&:=&{\alpha_0\over a_0}\left({1\over r}+{2\alpha_0'\over\alpha_0}\right).
\end{eqnarray}
Note that $A$, $B$, $C$ do not commute with each other, while
$\partial/\partial t$ commutes with all of them because their
coefficients are independent of $t$.

In this compact notation it is easy to see that by row operations we
can reduce the system (\ref{operatorform}) to the upper diagonal form
\begin{equation}
\left(\begin{array}{cc}
{\partial\over\partial t} & A \\
0 & \left({\partial\over\partial t}\right)^2 +C-BA
\end{array}\right)
\left(\begin{array}{cc}
\phi \\ \psi
\end{array}\right)=0.
\end{equation}
Hence we have reduced the time-dependent problem to the single
integral-differential equation
\begin{equation}
\label{psiH}
-\psi_{,tt}=H\psi, 
\end{equation}
where
\begin{equation}
\label{psiwaveequation1}
H:=C-BA=hg{\cal I}-({L}-g{\cal I}{v})({L}+g{v}{\cal I}).
\end{equation}
The coefficients $g$, $h$, $v$ and $V$ commute with each other but not
with the operators $\partial/\partial r$ or ${\cal
  I}$. $\partial/\partial r$ and ${\cal I}$ do commute with each other
because of the boundary condition (\ref{psiboundary}). 

It is convenient to split the time evolution operator $H$ into a
kinematic and a gravitational part as
\begin{equation}
H=H_0+H_1,
\end{equation}
with 
\begin{eqnarray}
\label{H0}
H_0&:=&-L^2=-V{\partial\over\partial r}V{\partial\over\partial r}, \\
\label{H1}
H_1&:=&g\left(h+\left({\cal I}v^2g\right)\right){\cal I}
+g{\cal I}v{L}-{L}gv{\cal I} \\
\label{H1bis}
&=&\left[g\left(h+\left({\cal I}v^2g\right)\right)-(Lgv)\right]{\cal I}
+g\left({\cal I}v{L}-v{L}{\cal I}\right). \nonumber \\
\end{eqnarray}

Hence we can write (\ref{psiH}) as
\begin{equation}
\label{psiwaveequation2}
-\psi_{,tt}+V{\partial\over\partial r}V{\partial\over\partial r}\psi=H_1\psi.
\end{equation}
This form stresses that (\ref{psiwaveequation2}) is a stratified wave
equation, in the sense that for fixed $Q$ and $F$ we have a wave
equation in $(t,r)$, with coefficients also depending on $Q$ and
$F$, while different values of $Q$ and $F$ are also coupled through
the double integral ${\cal I}$ over $Q$ and $F$, which represents the
gravitational interactions between particles of different momenta at
the same spacetime point. Intuitively, the characteristic speeds of
(\ref{psiwaveequation2}) are $\pm {V}$ because any matter perturbation
is propagated simply at the velocity of its constituent particles.

We note in passing that in our notation the static perturbations are
solutions of 
\begin{equation}
B\phi=0, \qquad \psi=0.
\end{equation}
Given a solution $\psi$ of the master equation (\ref{psiH}),
$\phi$ in any genuinely nonstatic perturbation can be reconstructed
from $\psi$ as
\begin{equation}
\label{phifrompsi}
\phi=-\left({\partial\over\partial t}\right)^{-1}A\psi,
\end{equation}
and the metric perturbations are obtained by solving
(\ref{deltaalphaprimem}-\ref{deltaaprimem}), with (\ref{deltaadotm})
obeyed automatically.

\subsection{Preliminary classification of non-static perturbations} 
\label{section:classificationofmodes}

The right-hand side of (\ref{psiwaveequation2}) is
generated by perturbations at all values of $Q$ and $F$ at at given
spacetime point $(t,r)$, but because of the overall factor $k_{,Q}$ it
only acts on perturbations at those values of $Q$ and $F$ where
$k_{,Q}(Q,F)\ne0$, that is, where background matter is present.

\begin{figure}
\includegraphics[scale=0.8,
  angle=0]{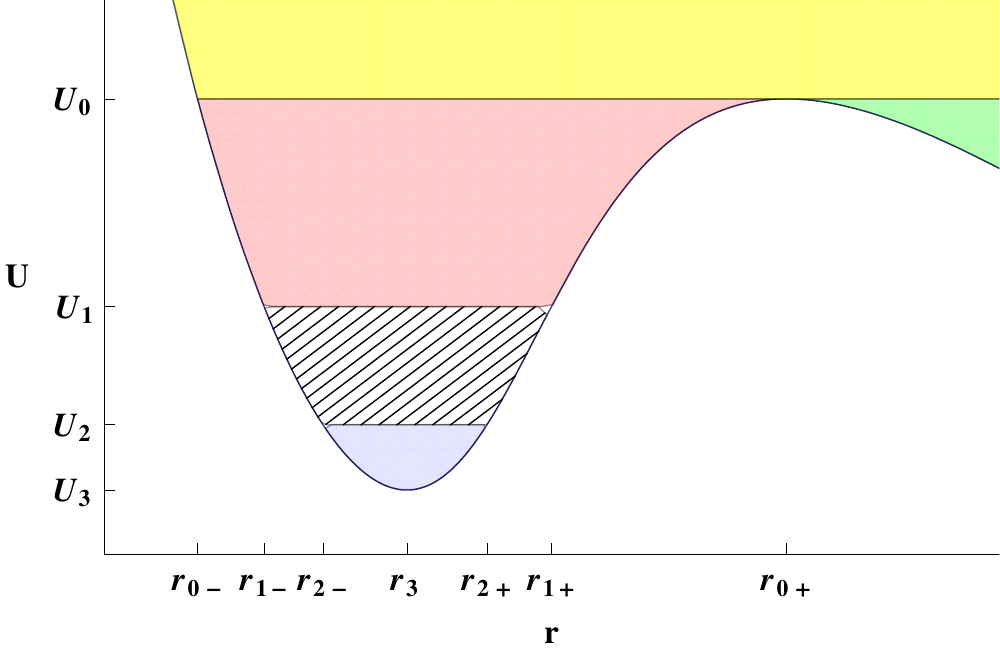}
\caption{Sketch of the potential $U(r)$ for massless particles,
  illustrating the classification of perturbations. The black hatching
  shows the region in $(U,r)$ space where particles are present in the
  background solution. Outer modes have support in the green region,
  top modes in the yellow region, middle modes in the red region,
  bottom modes in the blue region, and stellar modes in the hatched
  region only. (The top, middle and bottom modes also drive stellar
  modes.) This figure can also be interpreted as the cross-section
  $U(r,\infty)$ of the potential $U(r,F)$ in the massive case, compare
  the definition (\ref{Udef}) of $U$.}
\label{figure:potentialsketch}
\end{figure}

This suggests a preliminary classification of all perturbations into
\begin{itemize}
\item stellar modes, with support on the region in $(r,Q,F)$ space
  that lies inside the potential well and where also $k(Q,F)$ has support; 
\item bottom modes, with support inside the potential well and for
  values of $Q$ below the support of $k(Q,F)$;
\item middle modes, with support inside the potential well and for
  values of $Q$ above the support of $k(Q,F)$ but below the top of
  the potential well;
\item outer modes, with support outside the potential well and for
   values of $Q$ below the top of the potential well;
\item top modes, with values of $Q$ above the top of the potential
  well.
\end{itemize}
This classification of modes is illustrated in
Fig.~\ref{figure:potentialsketch} for the massless case, where all
values of $F$ experience the same effective potential
$U(r)$. Obviously, the middle modes do not exist if the potential well
is filled to the top ($U_1=U_0$), and the bottom modes do not exist if
it is filled to the bottom ($U_2=U_3$).

Particles contributing to the bottom, middle, top and outer modes move
freely in the effective potential $U(r,F)$ set by the static
background solution, without experiencing a gravitational
back-reaction. We may call them the trivial modes. Mathematically, the
trivial modes obey (\ref{psiwaveequation2}) in a region of $(Q,F)$
space where the right-hand side vanishes. In that region they can be
solved in closed form. To do this, we define the ``tortoise radius''
\begin{equation}
\label{sigmadef}
\sigma(r,Q,F):=\int^r {1\over {V}(r',Q,F)}\,dr'.  
\end{equation}
This gives us
\begin{equation}
L={V}\left.{\partial\over\partial r}\right|_Q
=\left.{\partial\over\partial \sigma}\right|_Q,
\end{equation}
and hence (\ref{psiwaveequation2}) becomes 
\begin{equation}
\label{psiwaveequation3}
-\psi_{,tt}+\psi_{,\sigma\sigma}=0
\end{equation}
for $\psi(t,\sigma,Q,F)$, 
subject to the Dirichlet boundary conditions
\begin{equation}
\label{psiboundarysigma}
\psi(t,\sigma_-(Q,F),Q,F)=\psi(t,\sigma_+(Q,F),Q,F)=0,
\end{equation}
where $\sigma_\pm$ are the left and right turning points given by $Q=U$.
Hence we can write down the general local solution of
(\ref{psiwaveequation2}) with $S=0$ in d'Alembert form as
\begin{equation}
\psi(t,r,Q,F)=\sum_\pm\psi_\pm\left[t\pm\sigma(r,Q,F),Q,F\right]
\end{equation}
subject to the boundary conditions (\ref{psiboundarysigma}).

With $g=0$ for the trivial modes, (\ref{phifrompsi}) becomes
$\phi_{,t}+\psi_{,\sigma}=0$, and hence
\begin{equation}
\phi(t,r,Q,F)=\sum_\pm\mp\psi_\pm\left[t\pm\sigma(r,Q,F),Q,F\right].
\end{equation}
We note also that $\psi+\phi$ must be non-negative for physical
trivial modes, as we cannot subtract particles from a vacuum region.

The trivial modes (except for the outer modes) induce stellar modes
through gravitational interactions, or mathematically through the
right-hand side of (\ref{psiwaveequation2}). This means that for a
complete solution of the problem, we need to solve for stellar modes
including an arbitrary driving term generated by the other modes, that
is
\begin{equation}
-\psi_{,tt}+V{\partial\over\partial r}V{\partial\over\partial r}\psi
=H_1\psi+H_1\psi_{\rm ext}
\end{equation}
where $\psi$ describes the stellar modes and $\psi_{\rm ext}$
is the sum of the top, middle and bottom modes. (The outer
modes do not couple to the stellar modes at all.)

\subsection{Massless case} 
\label{section:masslessperturbations}

In the massless case, we define the integrated matter perturbations
in the obvious way:
\begin{eqnarray}
{\bar\phi}(t,r,Q)&:=&\int_0^\infty \phi(t,r,Q,F)\,FdF, \\
{\bar\psi}(t,r,Q)&:=&\int_0^\infty \psi(t,r,Q,F)\,FdF.
\end{eqnarray}
The reduced perturbation equations with $m=0$ are obtained from the
ones in the general case by replacing $\phi$ and $\psi$ with
$\bar\phi$ and $\bar\psi$, $k(Q,F)$ and $k_{,Q}(Q,F)$ with $\bar k(Q)$
and $\bar k'(Q)$, and ${\cal I}$ with ${\bar{\cal I}}$. The
coefficients $g$, $v$, $V$ and $U$ all become independent of $F$. In
particular, all particles now move in the same effective potential
$U(r)$.

\section{Perturbation spectrum} 
\label{section:perturbationspectrum}

\subsection {Inner product} 
\label{section:innerproduct}

Starting from (\ref{H1}), we rewrite the gravitational part $H_1$ of
the Hamiltonian more explicitly as
\begin{eqnarray}
\label{H1full}
H_1&=&Xc_{1} {\cal I}
+X\left( c_{2} {\cal I} v^2{\partial\over\partial r}
-{\partial\over\partial r} c_{2}v^2 {\cal I} \right).
\end{eqnarray}
where we have defined the shorthands
\begin{eqnarray}
\label{Xdef}
X(Q,F,r)&:=&Qk_{,Q}V, \\
c_0(r)&:=&{8\pi^2 a_0^2\over r\alpha_0^2}, \\
c_{1}(r)&:=&{8\pi^2a_0\over r\alpha_0}
\left({1\over r}+{2\alpha_0'\over \alpha_0}+c_{3}\right),\\
c_{2}(r)&:=&{8\pi^2a_0\over r\alpha_0}, \\
c_{3}(r)&:=& {8\pi^2a_0^2\over r\alpha_0^2}{\cal I} {v}^3 \,Q\,k_{,Q}.
\end{eqnarray}
We have defined $c_0$ for later use: note that $g=c_0X$. For clarity
we have not written out function arguments, but recall that ${\cal I}$
acting on anything gives a function of $t$ and $r$ only, that
$k=k(Q,F)$, $v^2=1-U/Q$, $U=U(r,m^2/F)$, and hence that
$v=v(r,Q,m^2/F)$ and $V=V(r,Q,m^2/F)$.

We now construct an inner product $\langle\psi_1,\psi_2\rangle$ with
respect to which $H$ is a symmetric operator.  Consider first
solutions $\psi_1$, $\psi_2$ with support only where $k_{,Q}(Q,F)=0$, that
is the modes we have characterised as trivial. Then only $H_0\psi$ is
non-vanishing, and the perturbed Vlasov equation reduces to the free
wave equation (\ref{psiwaveequation3}). The obvious inner product is
therefore the well-known one for the free wave equation, that is
\begin{equation}
\label{innerproduct0}
\langle\psi_1,\psi_2\rangle:= \int_0^\infty\!\!\!\!\int_0^\infty\!\!
\left(\int_{\sigma_-(Q,F)}^{\sigma_+(Q,F)}
\psi_1\psi_2\,d\sigma
\right)\mu(Q,F) \,dQ\,FdF,
\end{equation}
where $\mu(Q,F)>0$ is an arbitray weight. Expressing this in terms of
$r$, we have
\begin{equation}
\label{innerproduct1}
\langle\psi_1,\psi_2\rangle=
\int_0^\infty\!\!\!\!\int_0^\infty\!\!\! 
\left(\int_{r_-(Q,F)}^{r_+(Q,F)} {\psi_1\psi_2\,dr\over V(r,Q,F)}\right)\mu(Q,F) 
\,dQ\,FdF.
\end{equation}

If we interchange the integration over $r$ with the double integration over
$Q$ and $F$, we obtain
\begin{eqnarray}
\label{innerproduct2}
\langle\psi_1,\psi_2\rangle&=&\int_0^\infty\!\!\!\!\int_0^\infty\!\!\!
\left(\int_{U(r,F)}^\infty {\mu(Q,F)\psi_1\psi_2 \,dQ\over
  V(r,Q,F)}\right)FdF\,dr \nonumber \\
&=&\int_0^\infty\!\!\left({\cal I}{\mu \psi_1\psi_2\over
  V}\right)dr
\label{innerproduct3}
\end{eqnarray}

Integration by parts in $r$ in (\ref{innerproduct1}), and then
transforming to (\ref{innerproduct3}), then gives 
\begin{equation}
\langle\psi_1,H_0\psi_2\rangle =\int_0^\infty\!\!\! \left({\cal I}\mu V
\psi_{1,r}\psi_{2,r}\right)\,dr
=\langle V\psi_{1,r},V\psi_{2,r}\rangle,
\label{innerproductfree}
\end{equation}
which is explicitly symmetric in $\psi_1$ and $\psi_2$.

Consider now solutions $\psi_1$, $\psi_2$ with support only where
$k_{,Q}(Q,F)\ne 0$, that is stellar modes. Assuming further that
$k_{,Q}<0$ wherever $k\ne 0$ (as will be the case for our examples),
we must then make the choice
\begin{equation}
\label{muchoice}
\mu(Q,F)=-{1\over Qk_{,Q}}
\end{equation}
(up to a positive constant factor), that is
\begin{eqnarray}
\label{innerproduct1a}
\langle\psi_1,\psi_2\rangle&=& -\int_0^\infty\!\!\! \left({\cal I}{\psi_1\psi_2\over
  X}\right)\,dr \\
\label{innerproduct1b}
&=&-\int_0^\infty\!\!\!\!\int_0^\infty\!\!\!\left(\int_{r_-}^{r_+}{\psi_1\psi_2\over
  X}\,dr\right)\,dQ\,FdF,
\end{eqnarray}
with $X$ defined in (\ref{Xdef}). 
We then find from (\ref{innerproductfree}), (\ref{innerproduct1a}) and
(\ref{H1full}) that
\begin{eqnarray}
\label{psiH0psi}
\langle\psi_1,H_0\psi_2\rangle &=&
-\int_0^\infty \left({\cal I}{V^2 \psi_{1,r}\psi_{2,r}\over X}\right)dr, \\
\label{psiH1psi}
\langle\psi_1,H_1\psi_2\rangle &=&
-\int_0^\infty c_{1}({\cal I}\psi_1)({\cal I}\psi_2) \,dr \nonumber \\
&&-\int_0^\infty 
c_{2}\bigl[({\cal I}\psi_1)({\cal I}v^2\psi_{2,r})\nonumber \\
&& \qquad +({\cal I}\psi_2)({\cal I}v^2\psi_{1,r})\big]\,dr, 
\end{eqnarray}
where for the last term we have used integration by parts in $r$.

We now define $\mu(Q,F)>0$ on all of phase space by (\ref{muchoice}) on
the support of $k(Q,F)$, and an arbitrary positive $\mu$, for example
$\mu=1$, everywhere else. 

\subsection{Quadratic form of the Hamiltonian} 
\label{section:quadraticform}

Consider the operator
\begin{equation}
\qquad {P}:={X{\cal I}\over ({\cal I}X)}.
\end{equation}
It is easy to see that ${P}$ is a projection operator, in the sense
that
\begin{equation}
{P}^2={P}.
\end{equation}
We also have
\begin{equation}
\langle \psi_1,{P}\psi_2\rangle=\int dr\, {\cal I}{\psi_1\over
  X}{X({\cal I}\psi_2)\over ({\cal I}X)}
=\int dr\, {({\cal I}\psi_1)({\cal I}\psi_2)\over({\cal I}X)},
\end{equation}
and so ${P}$ is symmetric,
\begin{equation}
{P}^\dagger={P}.
\end{equation}
As a projection operator, ${P}$ can only have eigenvalues $0$ and
$1$. The corresponding eigenspaces $\hat{\Bbb V}$ and $\bar{\Bbb V}$
are orthogonal, as
\begin{equation}
\langle (1-{P})\psi_1,{P}\psi_2\rangle
=\langle ({P}-{P}^2)\psi_1,\psi_2\rangle=0
\end{equation}
for any two vectors $\psi_1$, $\psi_2$.

Hence every normalisable vector $\psi$ can be uniquely split into
two vectors, one from each eigenspace, that is
\begin{equation}
\psi=\hat\psi+\bar\psi, \qquad \bar\psi:=P\psi,
\qquad \hat\psi:=\psi-\bar\psi,
\end{equation}
with ${\cal I}\hat\psi=0$. We write this statement as
\begin{equation}
\label{BbbVsplit}
{\Bbb V}=\hat {\Bbb V}\oplus \bar {\Bbb V}, \qquad 
\langle\hat {\Bbb V},\bar {\Bbb V}\rangle=0.
\end{equation}

From (\ref{psiH1psi}) we see that
\begin{equation}
\langle\hat {\Bbb V},H_1\hat {\Bbb V}\rangle=0,
\end{equation}
while all other matrix elements of $H_1$ and all of $H_0$ are
non-trivial.

Using integration by parts in $r$ in (\ref{innerproduct1b}) and the
boundary conditions
\begin{equation}
\psi(t,r_{\pm}(Q,F),Q,F)=0, 
\end{equation}
we have
\begin{eqnarray}
&& \langle \psi_1,L\psi_2\rangle 
=-\int FdF \int dQ {1\over Qk_{,Q}} \int_{r_-}^{r_+}dr\,
  \psi_1\psi_{2,r} \nonumber \\ 
&=&\int FdF \int dQ {1\over Qk_{,Q} }\int_{r_-}^{r_+}dr\, 
 \psi_{1,r}\psi_2 = \langle -L\psi_1,\psi_2\rangle, \nonumber \\
\end{eqnarray}
and so $L$ is antisymmetric, 
\begin{equation}
L^\dagger=-L. 
\end{equation}

Similarly, if we define
\begin{equation}
K:=gv{\cal I}=c_0Xv{\cal I}, 
\end{equation}
we have 
\begin{equation}
\langle \psi_1,c_0Xv{\cal I}\psi_2\rangle 
=-\int\! dr\,c_0\left({\cal I}v\psi_1\right)\left({\cal I}\psi_2\right) 
=\langle c_0X{\cal I}v\psi_1,\psi_2\rangle,
\end{equation}
and so
\begin{equation}
K^\dagger=c_0X{\cal I}v=g{\cal I}v.
\end{equation}
Hence we have
\begin{equation}
A=L+K, \qquad B=L-K^\dagger=-A^\dagger.
\end{equation}

We can also write $C$ as
\begin{equation} 
C=hg{\cal I}=hc_0X{\cal I}
=hc_0({\cal I}X){X{\cal I}\over({\cal I}X)}=-c_4P,  
\end{equation}
where we have defined the shorthand coefficient
\begin{equation}
c_4(r):=-c_0(r)h(r)({\cal I}X).
\end{equation}
Recall that we have assumed that $X\le 0$, and hence $({\cal I}X)\le
0$. Evidently $c_0>0$, and from (\ref{staticalpham}) we see that
$h>0$, so we have $c_4\ge 0$. From 
\begin{equation}
H=A^\dagger A-c_4P,
\end{equation}
we then obtain a (negative) lower bound on the spectrum of $H$, namely
$-\max_r c_4$. This is not sharp as $A\bar {\Bbb V}\ne 0$.

Using also that $P^2=P=P^\dagger$ and that
$P$ commutes with multiplication by any function of $r$, we have
\begin{equation}
C=-D^2, \qquad D:=\sqrt{c_4}P, \qquad D^\dagger=D.
\end{equation}
We can  therefore write the Hamiltonian also as the difference of two squares,
\begin{equation}
H=A^\dagger A-D^\dagger D.
\end{equation}

\subsection{Ritz method} 
\label{section:Ritzmethod}

We briefly review the Ritz method to establish notation. Given a
Hilbert space ${\Bbb V}$ with inner product
$\langle\cdot,\cdot\rangle$ and an operator $H$ that is self-adjoint
in ${\Bbb V}$, the method finds approximate eigenfunctions and
eigenvalues of $H$ in the span of a finite set of functions $e_i\in
{\Bbb V}$, $i=1\dots N$. (In contrast to the usual application in
quantum mechanics, our ${\Bbb V}$ is a real vector space.) The $e_i$
are assumed to have finite norm under the inner product, but need not be
orthogonal.

We try to find approximate eigenvectors $\psi$ of $H$ by determining
the coefficients $c^i$ in the ansatz
\begin{equation}
\label{finitebasis}
\psi=\sum_{i=1}^Nc^ie_i.
\end{equation}
Our notion of ``approximate eigenvector'' is defined relative to the
function set $\{e_i\}$, that is by
\begin{equation}
\label{approximateeigenvector}
\langle e_i,(H-\lambda)\psi\rangle = 0
\end{equation}
for $i=1\dots N$. We define the matrices
\begin{equation}
S_{ij}:=\langle e_i,e_j\rangle, \quad H_{ij}:=\langle e_i,H e_j\rangle.
\end{equation}
Then (\ref{approximateeigenvector}) is equivalent to
\begin{equation}
\label{generalisedeigenvectors}
\sum_{j=1}^N(H_{ij}-\lambda S_{ij})c^j=0
\end{equation}
for $i=1\dots N$. As $S_{ij}$ and $H_{ij}$ are real symmetric
matrices, the (approximate) eigenvalues $\lambda$ are real and the
corresponding (approximate) eigenvectors $\psi$ of the form
(\ref{finitebasis}) are orthogonal for different $\lambda$ (as must of
course be the case for the exact eigenvalues and eigenvectors of $H$).

\subsection{The space of test functions $\psi$} 
\label{section:functionspace}

We now attempt to restrict the real vector space ${\Bbb V}$ of
test functions $e_i$ in which we look for eigenfunctions of $H$. To
start with, we require functions in ${\Bbb V}$ to have a finite norm
under $\langle\cdot,\cdot\rangle$.

Starting from (\ref{H1bis}), we can write $H_1$ as 
\begin{equation}
\label{H1bisfull}
H_1=Xc_{2}\left[\left(c_{5}+c_{6}v^2+c_{7}{m^2\over
    FQ}\right){\cal I}+
{\cal I}v^2{\partial\over\partial r}
-v^2{\cal I}{\partial\over\partial r}\right],
\end{equation}
where we have defined the new shorthand coefficients
\begin{eqnarray}
c_{5}(r)&:=&{4\alpha_0'\over\alpha_0}-{1\over r}+c_{3} ,\\
c_{6}(r)&:=&-{\alpha_0'\over\alpha_0}-{a_0'\over a_0'}+{3\over r},\\
c_{7}(r)&:=&{2\alpha_0^2\over r}.
\end{eqnarray}
Note that 
\begin{equation}
H_1\psi=\left(\psi_1+\psi_2v^2+\psi_3{m^2\over
  FQ}\right)X(r,Q,F).
\end{equation}
The functions of one variable $\psi_1(r)$, $\psi_2(r)$ and $\psi_3(r)$ are
given by integrals of $\psi$, but we do not need their
explicit form for our argument.

Similarly, we can write $H_0\psi$ as
\begin{equation}
H_0\psi={\alpha_0^2\over a_0^2}
v^2\psi_{,rr}+\left(c_8+c_9v^2+c_{10}{m^2\over
  FQ}\right)\psi_{,r},
\end{equation}
where again the explicit form of the coefficients $c_8(r)$, $c_9(r)$
and $c_{10}(r)$ does not matter for the following argument.

Because $z=vZ$, with $Z$ an even function of $z$ by
(\ref{Zdefmassive}), and because we assume that $\psi(t,r,z,F)$ is
even in $z$, $\psi(t,r,Q,F)$ cannot be completely regular at the
boundary $Q=U$ where $v=0$. The closest to smoothness we can get is to
consider $\psi$ of the form
\begin{equation}
\label{psivchi}
\psi=\psi_r:=v(r,Q,F)\times\hbox{smooth}(t,r,Q,F),
\end{equation}
where the second factor is smooth in particular at $Q=U(r,F)$ and at
$Q=U_1(F)$. If $k_{,Q}(Q,F)$ is also smooth, in particular at the
boundary $Q=U_1(F)$ of its support, then equivalently we can consider
$\psi$ of the form
\begin{equation}
\label{psiXchi}
\psi=\psi_s:=X(r,Q,F)\times\hbox{smooth}(t,r,Q,F).
\end{equation}
From (\ref{H0}) and (\ref{H1bisfull}) we see that for smooth $k_{,Q}$,
$H$ maps functions of the form (\ref{psivchi}) into functions of the
same form. Hence if $Qk_{,Q}$ and therefore $X$ is smooth, we can
consistently restrict ${\Bbb V}$ to functions of the form
(\ref{psivchi}) or equivalently (\ref{psiXchi}).

However, we also want to consider background solutions where $k_{,Q}$
is not smooth at the boundary $Q=U_1$, and so $X$ is not smooth. The
key example of this is the class of critical solutions with massless
particles conjectured in Paper~I, which is characterised by $U_1=U_0$,
with $\bar k(Q)\sim (U_0-Q)^{1\over 2}$ for $Q\lesssim U_0$. Should we
now use the ansatz (\ref{psivchi}) or (\ref{psiXchi}), or a sum of
both? 

We note that $H_0$ 
maps each of (\ref{psivchi}) and (\ref{psiXchi}) into a function of the 
same form, whereas $H_1$ maps both to a function of the form
(\ref{psiXchi}). If we try the superposition
\begin{equation}
\label{psirs}
\psi=\psi_r+\psi_s,
\end{equation}
we find
\begin{equation}
\left(\begin{array}{cc}
(H\psi)_r \\ (H\psi)_s
\end{array}\right)=
\left(\begin{array}{cc}
H_0 & 0 \\
H_1 & H
\end{array}\right)
\left(\begin{array}{cc}
\psi_r \\ \psi_s
\end{array}\right),
\end{equation}
The eigenvalue problem $(H-\lambda)\psi=0$ then becomes
\begin{eqnarray}
(H_0-\lambda)\psi_r&=&0, \\
(H-\lambda)\psi_s&=&-H_1\psi_r.
\end{eqnarray}
Hence we can consistently assume that $\psi_r=0$, that is we can
restrict to the ansatz (\ref{psiXchi}). If we allow for the full
ansatz (\ref{psirs}), then $\psi_r$ behaves like a trivial mode,
driving a particular integral contribution to $\psi_s$. Hence we can
neglect $\psi_r$ when we are interested only in the spectrum of pure
stellar modes.

\subsection{Numerical examples of the Ritz method for massless particles}
\label{section:numericalexamples}

A set of basis functions of compact support $e_{mnp}(r,Q,F)$ naturally
has a triple discrete index $mnp$ corresponding to the index $i$ we used in the
general discussion of the Ritz method.  We focus here on the massless
case, where we can work directly with integrated modes
$\bar\psi(r,Q)$, and the integrations in $H$ and
$\langle\cdot,\cdot\rangle$ are the integrals $\bar{\cal I}$ over $Q$
only. The basis functions $e_{mn}(r,Q)$ then only carry two indices.

Within the massless case, we further focus on backgrounds
with the integrated Vlasov distribution given by ``Ansatz~1'' of
\cite{paperI}, or 
\begin{equation}
\label{backgroundansatz1}
\bar k(Q)\propto Q^{-(k+2)}(U_1-Q)^k_+,
\end{equation}
where $k\ge -1$ and $U_1$ are constant parameters, and the notation
$(\dots)_+^k$ stands for $\theta(\dots)(\dots)^k$. We then have
\begin{equation}
Q\bar k'(Q)\propto Q^{-(k+2)}(U_1-Q)_+^{k-1}[(k+2)U_1-2Q].
\end{equation}

This motivates the perturbation ansatz
\begin{eqnarray}
e_{mn}(r,Q)&:=&N_{mn}\,(U_1-Q)_+^m\,(r-r_3)^n\,\sqrt{Q-U(r)}
\nonumber \\ &&
[(k+2)U_1-2Q]\,
Q^{{l\over 2}-{k\over 2}-{5\over 4}}.
\end{eqnarray}
Here the constant factor $N_{mn}$ is a normalisation constant chosen
so that $\langle e_{mn},e_{mn}\rangle=1$. The next two factors carry
the basis indices $m$ and $n$, which we choose to be nonnegative
integers. The implied factor $\theta(U_1-Q)$ restricts us to stellar
modes (where $\psi$ can have either sign because there is a background
particle distribution from which we can subtract an infinitesimal
amount). We have chosen the integer powers of $U_1-Q$ and of $r-r_3$
as an ad-hoc basis of smooth functions of $Q$ and $r$ on an irregularly
shaped domain. We have chosen $r-r_3$ as this always changes sign on
the interval $[r_-(Q),r_+(Q)]$. The last factor on the first line
makes $e_{mn}$ odd in $z$, as previously discussed.

The two factors on the second line have been chosen merely for
convenience. Putting the factor $(k+2)U_1-2Q$ into our ansatz for
$e_{mn}$ either eliminates this same factor in the integrals over $Q$
in (\ref{innerproduct1a}), (\ref{psiH0psi}) and (\ref{psiH1psi}), or
at least puts it into the numerator, where it can be split into linear
factors already present and so does not make the integrand more
complicated. This leaves us with a sum of integrals of the form
\begin{equation}
\int_U^{U_1}Q^a(Q-U)^b(U_1-Q)^c\,dQ,
\end{equation}
where $b=-1/2$, $1/2$ or $3/2$, which can be evaluated in closed form
as hypergeometric functions of $1-U/U_1$. The final factor in the
ansatz allows us to set the power $a$ to a convenient, for example
integer, value by the corresponding choice of $l$. The integral over
$Q$ in the inner product (\ref{innerproduct3}) converges at $U=Q_1$
only if $m_1+m_2>k-2$, and so we must have $m>k/2-1$.

Note that the factor $(k+2)U_1-2Q$ is automically positive definite
only for $k>0$. For $k=0$ it can and should be removed from the
ansatz, as it would just increase $m$ by one. For $k<0$ our ansatz
with this factor works only for $(k+2)U_1>2U_3$ [which must be
verified numerically by finding $U(r)$]. 

We compute $S_{ij}$ and $H_{ij}$ by symbolic integration over $Q$
followed by numerical integration over $r$. We then solve
(\ref{generalisedeigenvectors}) directly. As in
\cite{AkbarianChoptuik} and Paper~I, we fix an arbitrary overall scale
in solutions of the massless Einstein-Vlasov system by setting the
total mass of the background solution to $1$.

As a first example, we take the background with $k=1$ and
$U_1=U_0=1/27$, the lip of the effective potential. For any positive
integer $k$, $\bar k'(Q)$ and hence $X$ is smooth at $Q=u_1$, and we
can consistently assume $m=0,1,2,\dots,M$ and $n=0,1,2,\dots,N$. We
set $l=11/2$, which simplifies the two $Q$-integrals in $S_{ij}$ and
$H_{0ij}$ somewhat, and reduces the two $Q$-integrals in $H_{1ij}$ to
polynomials.

As a second example, we take $k=1/2$ with $U_1=U_0=1/27$. This ansatz
seems to agree well with the critical solution observed in
\cite{AkbarianChoptuik}. We set $l=5$. All four integrals of $e_{mn}$
over $Q$ then reduce to polynomials of $U_1$ and $U(r)$. As discussed
above, we then choose $m=-1/2,1/2,3/2,\dots,M-1/2$, and
$n=0,1,2,\dots,N$ as before. In either example, the size of the basis,
and hence the number of approximate eigenvalues obtained, is $(M+1)(N+1)$.

\begin{figure}
\includegraphics[scale=0.6, angle=0]{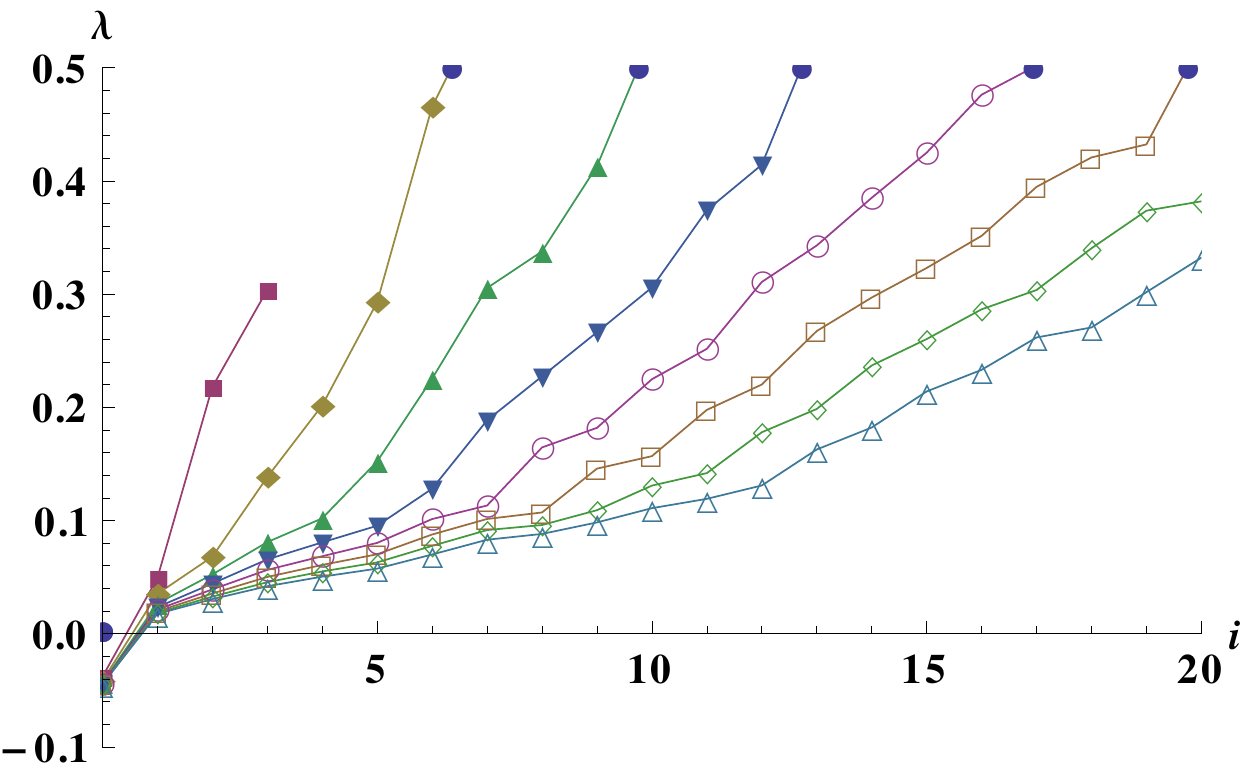} 
\includegraphics[scale=0.6, angle=0]{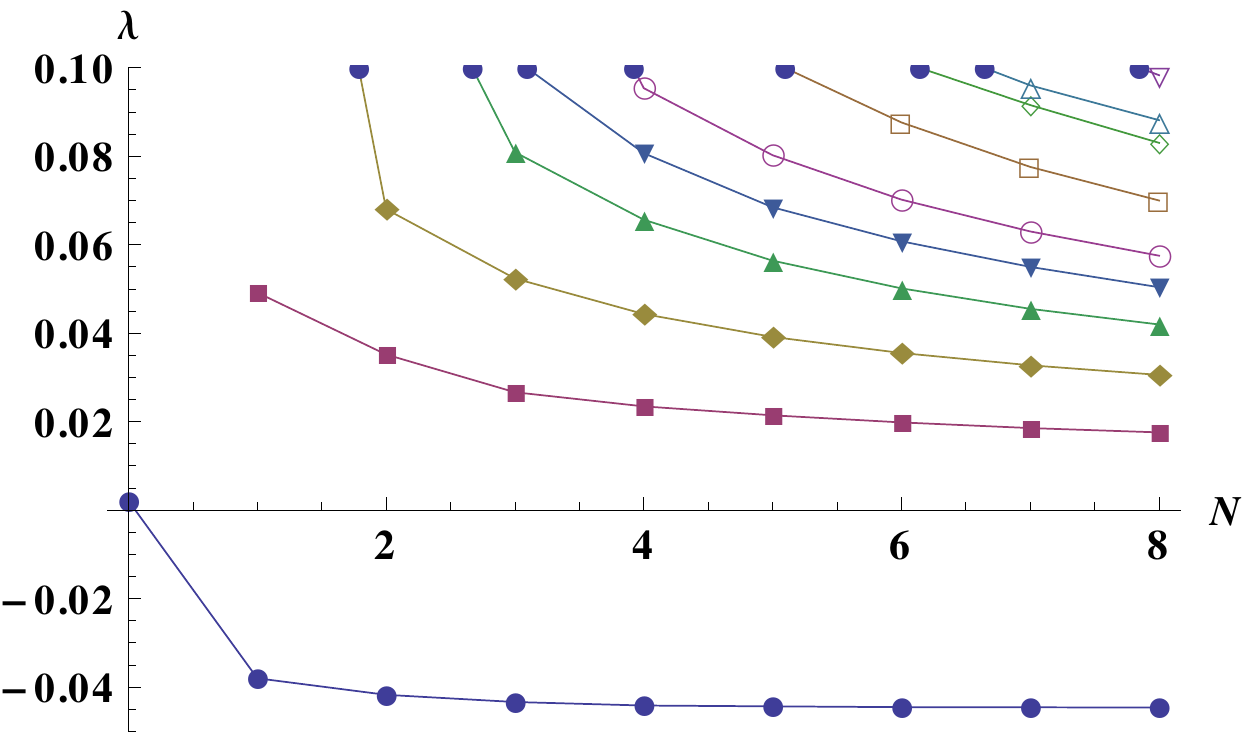} 
\caption{The eigenvalues $\lambda$ of (\ref{generalisedeigenvectors}),
  for the background given by the ansatz (\ref{backgroundansatz1})
  with $k=1$ and $U_1=U_0$, for different basis sizes
  $M=N=0,1,\dots,8$. In both plots, the vertical axis is $\lambda$. In
  the upper plot, the horizontal axis is the number $i$ of the
  eigenvalue, starting from $0$, and the different graphs correspond
  to the different resolutions. In the lower plot, the
  horizontal axis is resolution indicated by $N$, with $M=N$, and the
  different graphs show different eigenvalues.}
\label{figure:k1lambda}
\end{figure}

\begin{figure}
\includegraphics[scale=0.6, angle=0]{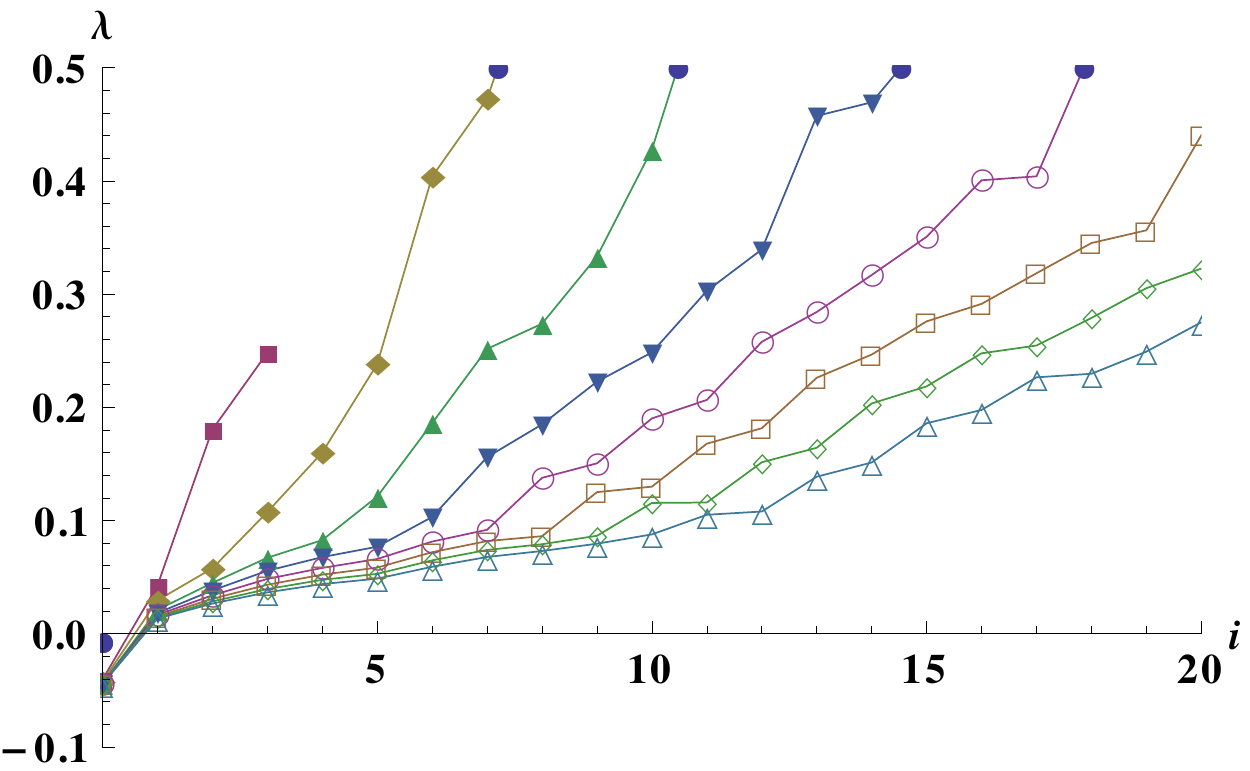} 
\includegraphics[scale=0.6, angle=0]{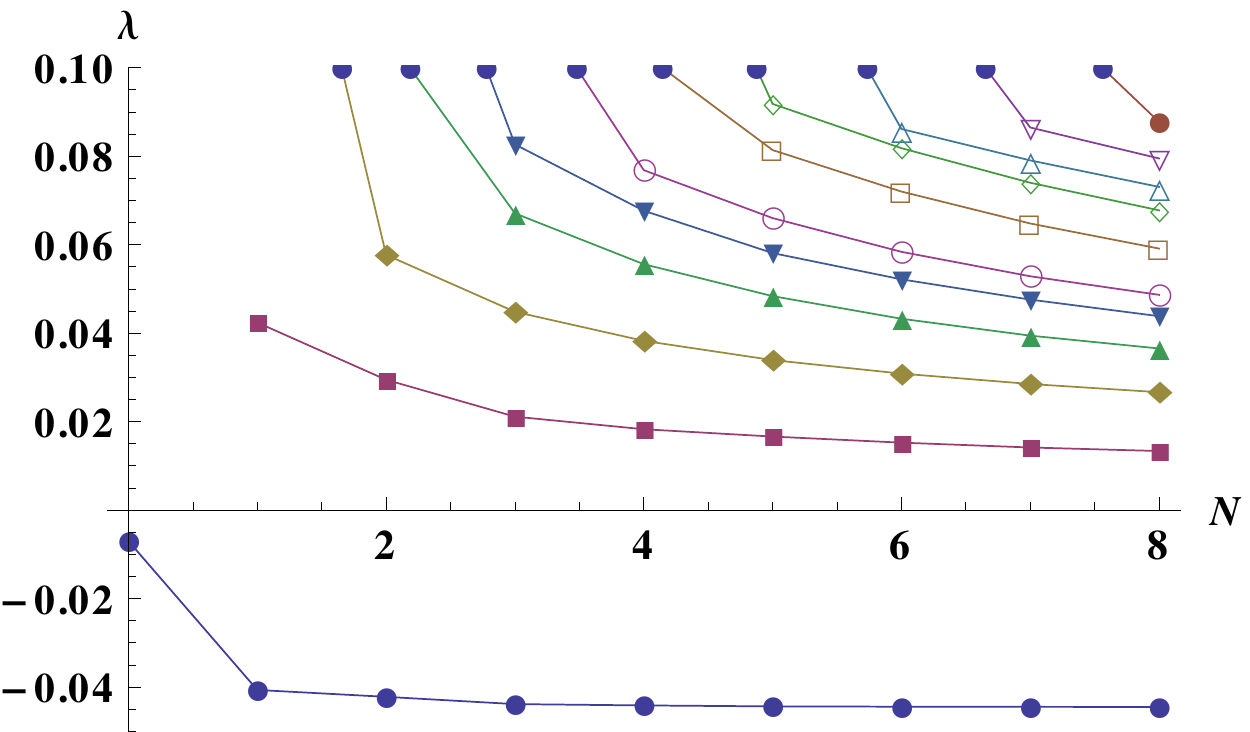} 
\caption{As before, but now for $k=1/2$.}
\label{figure:k12bislambda}
\end{figure}

In both examples, we have carried out the numerical integrations in
$r$ for $M=N=8$. The results are similar in both examples. At all
basis sizes up to and including this one, we find precisely one
negative eigenvalue of $H$, as expected from the results of
\cite{AkbarianChoptuik}. Its value is $\lambda_0\simeq -0.045$ in both
examples. 

The approximate eigenvalues for different basis sizes are shown in
Figs.~\ref{figure:k1lambda} and \ref{figure:k12bislambda}. The lowest few
eigenvalues appear to be converging with resolution to distinct values,
providing evidence for the consistency of our method and a discrete
spectrum.

The components $c^{mn}$ of the lowest eigenvector $\psi_0$ also seem
to converge with $N$, but to diverge with $M$. The reason may be that
the simple powers of $r-r_3$ and $Q-U_1$ are not good basis
functions. The corresponding eigenfunctions appear visually to be
smooth and converging with $M=N$ up to $M=N=7$. (In the $k=1/2$
example, $\psi_0$ diverges as expected, and this statement refers to
$\sqrt{U_0-Q}\,\psi_0$, which is finite.) For the largest basis
$M=N=8$, $\psi_0$ becomes noisy. Hence there is no point in increasing
$M$ or $N$ beyond $M=N=8$ with our limited accuracy.

If we write the dynamics of the time-dependent perturbations as
$-\psi_{,tt}=H\psi$, and $\lambda_0$ is the single negative eigenvalue
of $H$, then the single unstable mode has time-dependence
$\exp(\sqrt{-\lambda_0}t)$. This must correspond to
$\exp(\tau/\sigma)$ in the notation of \cite{AkbarianChoptuik}, where
$\tau$ is the proper time at the centre. $\tau$ is related to our
coordinate time $t$ (proper time at infinity) by $d\tau=\alpha_c\,dt$,
where $\alpha_c$ is the lapse at the centre, and hence we have
\begin{equation}
\label{sigmafromlambda0}
\sigma = {\alpha_c\over\sqrt{-\lambda_0}}.
\end{equation}
With $\alpha_c\simeq0.2968$ and $\lambda_0\simeq-0.0446$ for the
$k=1$ background solution, and $\alpha_c\simeq0.2874$ and
$\lambda_0\simeq-0.0445$ for $k=1/2$, the formula (\ref{sigmafromlambda0})
gives $\sigma\simeq 1.41$ and $1.36$, respectively, both compatible with the
range $\sigma\simeq1.43\pm 0.07$ given by
\cite{AkbarianChoptuik}. 

However, we have implemented our numerics using only standard ODE
solvers and nonlinear equations solvers (for solving the background
equations by shooting) and numerical integration and linear algebra
methods (for applying the Ritz method to the perturbations) in
Mathematica, and have not tried to estimate our numerical error or
optimise our methods.

\section{Conclusions} 
\label{section:conclusions}

Numerical time evolutions of the Einstein-Vlasov system in spherical
symmetry with massless particles \cite{AkbarianChoptuik} have
suggested the rather surprising conjecture that all static solutions
of this system are one-mode unstable, with the time evolutions
resulting in collapse for one sign of the initial amplitude of this
mode, and dispersion for the other. In the language of critical
phenomena in gravitational collapse, all static solutions are critical
solutions at the threshold of collapse. 

In Paper~I \cite{paperI} we have characterised all static solutions
with massive particles in terms of a single function of two variables
$k(Q,F)$. Here $F$ is essentially conserved angular momentum, and $Q$
essentially conserved energy per angular momentum, such that the orbit
of a {\em massless} particle of given $Q$ and $F$ depends on $Q$
alone. Correspondingly, we have the degeneracy that all distributions
$k(Q,F)$ of massless particles with the same $\bar k(Q):=\int
k(Q,F)\,FdF$ give rise to the same spacetime. 

In the current Paper~II we have reduced the perturbations of static
solutions (in spherical symmetry, with either massive or massless
particles) to a single master variable $\psi(t,r,Q,F)$, which obeys an
equation of motion of the form $-\psi_{,tt}=(H_0+H_1)\psi$. In the
massless case we have the same degeneracy for the perturbations as for
the background, that is the metric perturbations only depend on
$\bar\psi=\int\psi\,FdF$, but in contrast to the background
equations this does simplify the equations significantly. Hence we
have assumed $m\ge 0$ in most of this paper.

The kinetic part $H_0$ of $H$ is such that $\psi_{,tt}=H_0\psi$ is
simply a second-order wave equation with characteristic speeds $\pm
V(r,Q,F)$. There is no underlying wave equation in three space
dimensions here. Rather, the left and right-going waves correspond to
particles moving inwards and outwards in the background spacetime,
with $V$ their radial velocity.

By contrast, the gravitational part $H_1$ of $H$ vanishes in the
vacuum regions of the background solutions, and contains integrals
over $Q$ and $F$ (as well as first $r$-derivatives) representing the
gravitational pull of all the other particles represented by the
perturbation $\psi$.

Following a suggestion by Olivier Sarbach, we have identified an inner
product of perturbations $\psi$ which is positive definite for
suitable background solutions and with respect to which the operator
$H$ is symmetric. This additional mathematical structure allows us to find
approximate eigenvectors $\psi$ and eigenvalues $\lambda$ of $H$ using
the Ritz method. We have carried out the numerical procedure for two
representative background solutions with massless particles (see Paper~I
for a discussion of these solutions and their significance), and we
have found numerical evidence for a discrete spectrum of $\lambda$
with, for both backgrounds, a single negative eigenvalue with a value
compatible with that found by Akbarian and Choptuik \cite{AkbarianChoptuik}.

On the analytic side, we have characterised the space of functions
$\psi(r,Q,F)$ as ${\Bbb V}=\hat {\Bbb V}\oplus \bar {\Bbb V}$, where a
certain integral ${\cal I}\psi$ over $F$ and $Q$ vanishes for
functions in $\hat {\Bbb V}$, while $\bar {\Bbb V}$ consists of
functions of the form $f(r)X(r,Q,F)$ for a specific $X$ given
by the background solution. Hence $\hat {\Bbb V}$ is infinitely larger
than $\bar {\Bbb V}$. Unfortunately, eigenvectors $\psi$ of
$H$ cannot lie entirely in either subspace. We have also found that we
can write $H$ as the difference of two squares, $H=A^\dagger
A-D^\dagger D$, with $D=D^\dagger$ annihilating $\hat {\Bbb V}$. Unfortunately,
the commutators of $A^\dagger$, $A$ and $D$ are not simple, and so the
apparent analogy with the quantisation of the harmonic oscillator does
not seem to be helpful.

We had hoped that in bringing the perturbation equations into a
sufficiently simple form we could prove the conjecture of
\cite{AkbarianChoptuik} that every spherically symmetric static
solution with massless particles has precisely one unstable mode
and/or calculate its value in closed form. We had also hoped to be
able to show that some spherically symmetric static solutions with
massive particles are stable, as conjectured in
\cite{AndreassonRein2006}. We have not been able to do either, but
hope that our formulation of the problem will be of future use.


\acknowledgments

The author acknowledges financial support from Chalmers University of
Technology, and from the Erwin Schr\"odinger International Institute
for Mathematics of Physics during the workshop ``Geometric Transport
Equations in General Relativity''. He is grateful to H\aa kan
Andr\'easson for stimulating discussions when this project was begun,
and to workshop participants Olivier Sarbach for suggesting the Ritz
method and Gerhard Rein for pointing out \cite{DeJonghe}.



\end{document}